\documentclass[aps,showpacs,floatfix,twocolumn,pra]{revtex4}

\usepackage{epsfig,amsmath,graphicx,graphics,float}

\begin{document}

\title{Universal contact and collective excitations of a strongly
interacting Fermi gas}

\author{Yun Li}
\author{Sandro Stringari}
\affiliation{Dipartimento di Fisica, Universit\`{a} di Trento and
INO-CNR BEC Center, I-38123 Povo, Italy}

\begin{abstract}

We study the relationship between Tan's contact parameter and the
macroscopic dynamic properties of an ultracold trapped gas, such as
the frequencies of the collective oscillations and the propagation
of sound in one-dimensional (1D) configurations. We find that the
value of the contact, extracted from the most recent low-temperature
measurements of the equation of state near unitarity, reproduces
with accuracy the experimental values of the collective frequencies
of the radial breathing mode at the lowest temperatures. The
available experiment results for the 1D sound velocities near
unitarity are also investigated.

\end{abstract}


\maketitle

\section{Introduction}

Significant experimental and theoretical work has been devoted in
recent years to understand the universal properties of interacting
Fermi gases along the BEC-BCS crossover (for a review, see, for
example, \cite{Giorgini2008}). More recently, Tan has introduced a
new concept for investigating universality based on the so-called
contact parameter, which relates the short-range features of these
systems to their thermodynamic properties \cite{Tan2008,Werner2009}.
The universality of the Tan's relations has been proven in a series
of experiments based on the measurement of the molecular fraction
\cite{Partridge2005}, the momentum distribution, the RF
spectroscopic rate at high frequencies, the adiabatic sweep and
virial theorems \cite{Stewart2010}, the spin structure factor
\cite{Kuhnle2010} and the equation of state \cite{Navon2010}. The
temperature dependence of the contact parameter has been the object
of  recent theoretical \cite{Hu2011} and experimental
\cite{Kuhnle2011} work.

In this paper we discuss the relationship between Tan's contact
parameter and the frequencies of the collective oscillations of a
harmonically trapped Fermi gas near unitarity. We also investigate
the relationship with the sound velocity in highly elongated
configurations. The study of the collective oscillations along the
BEC-BCS crossover in terms of the contact parameter has been already
addressed in \cite{DelloStritto2010} where upper bounds to the
collective frequencies were calculated using a sum rule approach.
However, the sum rule method developed in this work significantly
overestimates the hydrodynamic frequencies of trapped Fermi gases,
being consequently ineffective for a useful quantitative comparison
with experimental data in the superfluid hydrodynamic regime of low
temperatures.

The present approach is based on a perturbative solution of the
hydrodynamic equations of superfluids near unitarity
\cite{Bulgac2005}. This allows for an exact analytic relationship
between Tan'contact parameter and the deviations of the frequencies
of the collective oscillations as well as of the one-dimensional
(1D) sound velocity from their values at unitarity. The high
precision achievable in the frequency measurements is in particular
expected to provide an alternative accurate determination of the
contact parameter and to further confirm the universality of this
physical quantity.

\section{Contact, equation of state and hydrodynamic equations}

We start from the following definition of the contact parameter,
based on the so-called adiabatic sweep theorem \cite{Tan2008}:
\begin{equation}
\left[\frac{dE}{d(1/a)}\right]_N = -\frac{\hbar^2 \mathcal{I}}{4\pi
M} ,\label{eq:adiabatic}
\end{equation}
where $E$ is the total energy of the system, $\mathcal{I}$ is the
contact parameter, $M$ is the atomic mass, and $a$ is the $s$-wave
scattering length. By using the local density approximation (LDA),
the total energy can be calculated as $E=\int d^3r(\epsilon+ n
V_{\text{ext}})$, where $V_{\text{ext}}$ is  the trapping potential
and $\epsilon$ is the energy density of a uniform gas. The
equilibrium profile, in the LDA, is determined by the equation
\begin{equation}
\mu(n)+V_{\text{ext}}=\bar{\mu} , \label{eq:LDA}
\end{equation}
where
\begin{equation}
\mu(n)=\frac{\partial \epsilon(n,a)}{\partial
n} \label{eq:mu}
\end{equation}
is the chemical potential of uniform matter, providing the equation
of state of the gas, while $\bar{\mu}$ is the chemical potential of
the trapped system, fixed by the normalization condition. The
derivative of the total energy with respect to $1/a$ in
Eq.(\ref{eq:adiabatic}) can then be conveniently written as
\begin{equation}
\left[\frac{dE}{d(1/a)}\right]_N=\int d^3r\, \left[\frac{\partial
\epsilon(n,a)}{\partial (1/a)}\right]_n ,\label{eq:adiabatic2}
\end{equation}
exploiting the link between the contact parameter and the equation
of state.

On the other hand the chemical potential of uniform matter $\mu(n)$
is a crucial ingredient of the hydrodynamic equations of
superfluids. At zero temperature, these equations actually read
\cite{Pitaevskii2003}:
\begin{equation}
\begin{aligned}
&\frac{\partial n}{\partial t}+\nabla\left(\mathbf{v} n\right)=0 ,\\
&\frac{\partial \mathbf{v}}{\partial
t}+\frac{1}{M}\nabla\left[\frac{1}{2}M \mathbf{v}^2 +\mu(n)+
V_{\text{ext}} \right]=0,
\end{aligned}\label{eq:HD_equation}
\end{equation}
so that an insightful question is to understand the link between the
contact parameter and the behavior of the collective modes emerging
from the solutions of Eq.(\ref{eq:HD_equation}). In the following we
will discuss the problem near unitarity, where we can expand the
chemical potential in the form \cite{Tan2008,Bulgac2005}
\begin{equation}
\begin{aligned}
\mu(n)=\frac{\hbar^2}{2M}\xi \left(3 \pi^2 n \right)^{2/3}
-\frac{2\hbar^2}{5M} \frac{\zeta}{a}\left(3\pi^2 n\right)^{1/3}
\end{aligned}\label{eq:mu_expand}
\end{equation}
with $\xi$ and $\zeta$ being two universal dimensionless parameters
accounting for the effects of the interactions. The first term in
(\ref{eq:mu_expand}) exhibits the same density dependence as the
ideal Fermi gas with the renormalization factor $\xi$. The second
term (first-order correction in $1/a$) is directly related to Tan's
parameter calculated at unitarity. Indeed, using
Eqs.(\ref{eq:adiabatic}-\ref{eq:adiabatic2}), one easily finds that
the contact, for a harmonically trapped system, is given by
\begin{equation}
\frac{\mathcal {I}}{N k_F} =\frac{512}{175}\zeta,
\label{eq:Tancontact}
\end{equation}
where we have defined the Fermi
wave vector $k_F=[3\pi^2 n(0)]^{1/3}$ depending on the density in
the center of the trap.

The small-amplitude oscillations of the gas can be studied by
solving the linearized hydrodynamic equations
(\ref{eq:HD_equation}),
\begin{equation}
-\omega^2\delta n=\frac{1}{M}\nabla\cdot\left\{n\nabla\left[
\frac{\partial \mu(n)}{\partial n}\delta n\right] \right\},
\label{eq:HD_equation_lin}
\end{equation}
where $\delta n$ is the amplitude of the density oscillations around
the equilibrium value $n$. At unitarity $\mu(n) \propto n^{2/3}$,
and the solutions of Eq.(\ref{eq:HD_equation_lin}), in the presence
of harmonic trapping with axial symmetry, exhibit the analytic form
\begin{equation}
\omega^2(\lambda)= \frac{\omega_\perp^2}{3}\left[\left(4\lambda^2+ 5
\right) \pm \sqrt{16\lambda^4-32\lambda^2 +25 } \,\right],
\label{eq:dispersion_law}
\end{equation}
with $\lambda=\omega_z/\omega_\perp$, holding for the lowest $m=0$
modes, and include the coupling between the monopole and quadrupole
oscillations caused by the non spherical shape of the potential
(here $m$ is the $z$ component of angular momentum carried by the
excitation). Notice that, remarkably, result
(\ref{eq:dispersion_law}) does not depend on the parameter $\xi$
characterizing the equation of state at unitarity. This relation for
the collective frequencies was actually first obtained in contexts
different from the unitary Fermi gas, such as the ideal Bose gas
above $T_c$ \cite{Kagan1997,Griffin1997} and the ideal Fermi gas
\cite{Amoruso1999} in the hydrodynamic regime. In both cases the
equation of state [$\mu(n,s)$ actually exhibits the same $n^{2/3}$
dependence for fixed entropy per particle $s$] and the hydrodynamic
equations yield the same dispersion law (\ref{eq:dispersion_law})
for the scaling solutions of coupled quadrupole monopole type in the
presence of harmonic trapping. The prediction
(\ref{eq:dispersion_law}) for the collective frequencies was checked
experimentally at unitarity providing a direct confirmation of the
universality exhibited by the unitary Fermi gas \cite{Kinast2004,
Altmeyer2007}. The same oscillations were investigated out of
unitarity along the BEC-BCS regime, confirming the predictions of
theory \cite{Stringari2004, Astrakharchik2005} and in particular the
fine details of the equation of state accounted for   by quantum
Monte Carlo simulations.

\section{Collective frequencies and sound velocity shifts near
unitarity}

In the following we calculate the deviations of the collective
frequencies from the unitary value (\ref{eq:dispersion_law}) holding
for small values of the dimensionless parameter $1/k_F a$. To this
purpose we solve the hydrodynamic equations using a perturbative
procedure, which is generally applicable to any equation of state
having the form $\mu(n)=\mu_0(n)+\mu_1(n)$, where
$\mu_1(n)\ll\mu_0(n)$ represents the first-order correction. The
density profile $n(\mathbf{r})=n_0(\mathbf{r})+n_1(\mathbf{r})$,
including the first-order correction, can be obtained starting from
the expansion of $\mu(n)$ around the zero-order ground-state density
profile $n_0(\mathbf{r})$ and using the equilibrium condition
(\ref{eq:LDA}) in LDA. This gives
\begin{equation}
n_1=\left[\delta\bar{\mu}-\mu_1(n_0)\right]\left/ \frac{\partial
\mu_0(n_0)}{\partial n_0}\right. ,\label{eq:n1_tot}
\end{equation}
where $\delta\bar{\mu}$ is the first-order correction to the
chemical potential $\bar{\mu}$. Solving the linearized hydrodynamic
equation (\ref{eq:HD_equation_lin}) perturbatively, we find the
following expression for the frequency shift:
\begin{equation}
\frac{\delta \omega}{\omega} = -\frac{\int d^3 r\left(\nabla^2
f_0^\ast\right) \left[n_1-n_0 \left(\partial n_1/\partial n_0\right)
\right] f_0}{2\omega^2 M\int d^3 r \,f_0^\ast \;\delta n_0}
\label{eq:delta_omega_BF}
\end{equation}
for the collective oscillations caused by the perturbation
$\mu_1(n)$ in the chemical potential. In
Eq.(\ref{eq:delta_omega_BF}) $f_0=[\partial \mu_0(n_0)/\partial
n_0]\delta n_0$ is the zero-order eigenfunction of
(\ref{eq:HD_equation_lin}), and $n_1$ is given by (\ref{eq:n1_tot}).
According to Eq.(\ref{eq:delta_omega_BF}), one always has $\delta
\omega=0$ for the surface modes satisfying the condition $\nabla^2
f_0 =0$, as expected, due to their independence of the equation of
state. For compression modes one instead expects a correction due to
the changes in the equation of state. In general, one can show that
the first term in (\ref{eq:n1_tot}) proportional to
$\delta\bar{\mu}$ gives no contribution to the frequency shift and
will be consequently neglected in the following \cite{Note_02}.

Result (\ref{eq:delta_omega_BF}) is valid for both Bose and Fermi
systems. In the case of weakly interacting Bose-Einstein condensed
gas, it allows for the calculation of the frequency shifts caused by
the Lee-Huang-Yang corrections in the equation of state. In this
case, the corresponding density correction, neglecting the term
proportional to $\delta\bar{\mu}$, can be written as $n_1=-32 (n_0
a)^{3/2}/3 \sqrt{\pi}$, and using Eq.(\ref{eq:delta_omega_BF}), one
finds the frequency shift of the compression mode as derived in
\cite{Pitaevskii1998}. For the Fermi gas near unitarity, we instead
employ the expansion of the equation of state (\ref{eq:mu_expand})
around the density profile $n_0(\mathbf{r})$ calculated at
unitarity. Ignoring also in this case the irrelevant term
proportional to $\delta\bar{\mu}$, we find
\begin{equation}
n(\mathbf{r})=n_0(\mathbf{r})+n_1(\mathbf{r})=n_0(\mathbf{r})
-\frac{3\beta}{2\alpha} n_0^{2/3} (\mathbf{r}),
\label{eq:density_tot}
\end{equation}
where $\alpha=\hbar^2 \xi(3 \pi^2 )^{2/3}/2M$, $\beta =-2\hbar^2
\zeta(3\pi^2)^{1/3}/(5Ma)$, and $n_0= [(\bar{\mu}_0- V_{\text{ext}})
/\alpha]^{3/2}$, with $\bar{\mu}_0$ being the chemical potential
evaluated for a trapped system at unitarity. After some
straightforward algebra, one finds the following expression for the
frequency shift of the collective oscillations near unitarity:
\begin{equation}
\frac{\delta \omega}{\omega}= \frac{\beta}{6\omega^2 M} \frac{\int
d^3 r\left(\nabla^2 f_0^\ast\right)n_0^{2/3} \,f_0}{\int d^3 r \,
f_0^\ast n_0^{1/3}\, f_0} .\label{eq:delta_omega_general}
\end{equation}
One can check that this result is equivalent to the result of
Eq.(24) and (25) in \cite{Bulgac2005} where a similar expansion was
carried out near unitarity.

\begin{figure}
\includegraphics[scale=0.38]{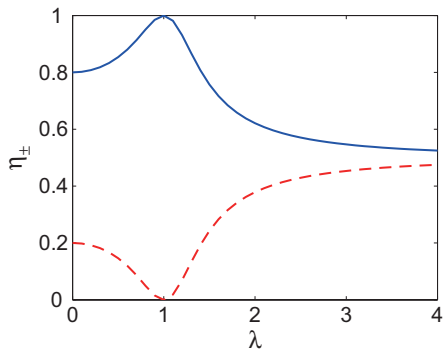}
\caption{(Color online) Functions $\eta_{+}$ (solid blue) and
$\eta_{-}$ (dashed red) relative to the higher and lower $m=0$ modes
as a function of the deformation parameter
$\lambda=\omega_z/\omega_{\perp}$.}\label{fig:eta}
\end{figure}

The eigenfunctions for the $m=0$ modes (\ref{eq:dispersion_law})
have the form $f_0 \sim a+br_\perp^2+cz^2$, where
\begin{equation}
\begin{aligned}
\frac{a}{b}&=-\frac{\bar{\mu}_0[4\lambda^2+3(\omega/\omega_\perp)^2
-10]}{6M\omega_z^2},\\
\frac{c}{b}&=\frac{3(\omega/\omega_\perp)^2-10}{2}.
\end{aligned}
\end{equation}
After some length but straightforward algebra, one finally obtains
the following result for the frequency shift:
\begin{equation}
\frac{\delta \omega}{\omega}= \left[\frac{128\zeta}{525\pi
\xi}\,\eta_{\pm}(\lambda) \right] \left(k_F a\right)^{-1} =
\left[\frac{\mathcal{I}/ N k^0_F}{12\pi \xi^{1/2}}
\,\eta_{\pm}(\lambda)\right] \left(k^0_F a\right)^{-1},
\label{eq:delta_omega}
\end{equation}
where
\begin{equation}
\eta_{\pm}(\lambda)= \frac{1}{2}\pm\frac{3}{ \phantom{\hat{1}}
2\sqrt{16 \lambda^4-32 \lambda^2+25} \phantom{\hat{1}}},
\end{equation}
with the index $\pm$ referring to the higher $(+)$ and lower $(-)$
frequencies of (\ref{eq:dispersion_law}). In the second equality of
(\ref{eq:delta_omega}), we have used relation (\ref{eq:Tancontact})
for the contact parameter calculated for an harmonically trapped
atomic cloud and we have expressed the Fermi momentum $k_F$ in terms
of the ideal Fermi gas wave vector $k^0_F=(24N)^{1/6}a_{ho}^{-1}$,
with $a_{ho}$ being the geometrical average of the harmonic
oscillator lengths. For the same total number of atoms, the density
$n(0)$ in the center of the trap for the unitary gas is $\xi^{-3/4}$
times larger than for the ideal gas, yielding $k_F=k^0_F\xi^{-1/4}$.

Eq.(\ref{eq:delta_omega}) represents the main result of the present
paper. It relates Tan's contact, a central quantity for the
universality relations holding in interacting systems, with the
low-energy macroscopic dynamics of the system, namely, the
frequencies of the collective oscillations. These equations can be
used either to predict theoretically the frequency shifts, once the
dimensionless parameters $\xi$ and $\zeta$ or Tan's contact are
known, or to determine experimental constraints on the value of
$\mathcal{I}$. In Fig.\ref{fig:eta} we plot $\eta_{\pm}$ as a
function of the deformation parameter $\lambda$. For a spherical
trap ($\lambda=1$) one obtains $\eta_+=1$ for the monopole mode and
$\eta_-= 0$, confirming, as already anticipated, that there is no
frequency shift for the surface quadrupole mode. When $\lambda \neq
1$, the two modes are coupled. In the limit of both spherical traps
and highly elongated traps ($\lambda \ll 1$) it reproduces the
results of \cite{Bulgac2005}.

\begin{figure}
\includegraphics[scale=0.32]{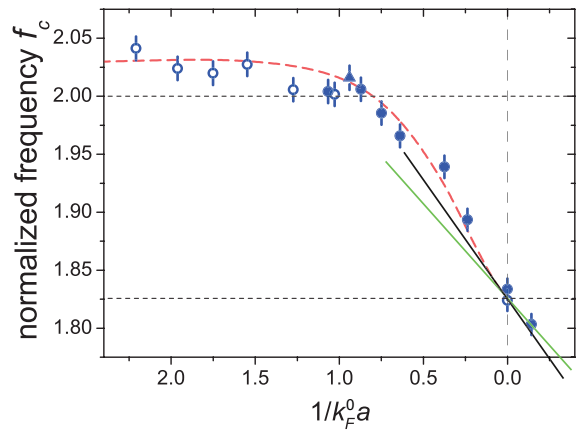}
\caption{(Color online) Frequency of the radial compression mode for
an elongated Fermi gas in the units of radial frequency
($\omega_\perp$) \cite{Altmeyer2007}. The deformation of the trap
$\lambda=0.038$. The dashed red curve refers to the equation of
state based on Monte Carlo simulations. Open and solid circles
correspond to experiment measurements. The black straight line shows
the slope of the frequency shifts around unitarity obtained from
Eq.(\ref{eq:delta_omega}) using $\xi=0.41$ and $\zeta=0.93$ as
extracted from the direct measurement of the equation of state
\cite{Navon2010} (corresponding to $\mathcal{I}/N k^0_F\simeq 3.4$).
The green (light gray) straight line instead corresponds to the
values $\xi=0.41$ and $\zeta =0.74$ yielding $\mathcal{I}/N
k^0_F\simeq 2.7$ (see text).} \label{fig:Grimm}
\end{figure}

In Fig.\ref{fig:Grimm} we show the prediction of
Eq.(\ref{eq:delta_omega}) in the trapping conditions of the
experiment of \cite{Altmeyer2007}, using the values $\xi=0.41$,
$\zeta=0.93$ extracted from the direct measurement of the equation
of state \cite{Navon2010} carried out at the lowest temperatures and
yielding the value $\mathcal{I}/Nk_F^0\simeq 3.4$ for the contact
parameter. These values differ by a few percent from the most recent
theoretical predictions based on Monte Carlo simulations at $T=0$
(see, for example, \cite{Gandolfi2011} and references therein). The
predicted slope $\delta\omega/\omega \sim 0.11/(k^0_Fa)$ (black
line) of the collective frequencies of the radial breathing mode at
unitarity turns out to be in very good agreement with experiments
($\delta \omega/\omega \sim 0.12/(k_F^0 a)$ \cite{Note_01}).

In order to appreciate the sensitivity of the slope to the choice of
the values of $\xi$ and $\zeta$, in Fig.\ref{fig:Grimm}, we also
show the predictions [green (light gray) line] for the frequency
shifts using the values $\xi=0.41$ and $\zeta=0.74$, yielding the
smaller value $\mathcal{I}/N k^0_F\simeq 2.7$ for the contact. This
value is closer to the measurement of the contact carried out in
\cite{Stewart2010} and \cite{Kuhnle2011} at slightly higher values
of temperature. The resulting slope $\delta \omega/\omega \sim
0.09/(k_F^0 a)$ provides a worse description of the experimental
data for the collective frequencies.

The collective oscillations discussed above represent the
discretized values of the usual sound waves described by
hydrodynamics. It is actually useful to calculate also the changes
of the sound velocity of a uniform sample near unitarity in terms of
the dimensionless parameters $\xi$ and $\zeta$ or, equivalently,
Tan's contact parameter. For bulk Fermi gases, one finds $\delta c
/c = -\zeta/(5\xi k_F a)$.

\begin{figure}
\includegraphics[scale=0.43]{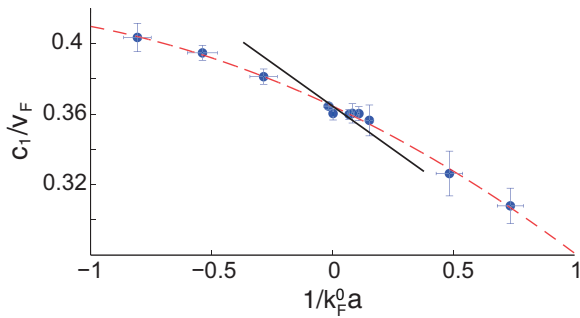}
\caption{(Color online) Normalized 1D sound velocity $c_1/v_F$ vs
the interaction parameter $1/k^0_Fa$ in the unitary regime, where
$v_F =\hbar k_F^0/M$ is the Fermi velocity for ideal Fermi gas and
$k_F^0 =k_F \xi^{1/4}$ is the corresponding Fermi wave vector.
Circles with error bars correspond to the experiment measurements
\cite{Joseph2007}. The dashed red curve is a quadratic fitting
\cite{Thomas_private}, and the black straight line is the slope of
$\delta c_1/c_1$ calculated from Eq.(\ref{eq:delta_c1}) using the
values $\xi=0.41$ and $\zeta=0.93$ extracted from
\cite{Navon2010}.}\label{fig:Thomas}
\end{figure}

In the case of cylindrical geometry with radial harmonic trapping
the sound velocities can be also calculated starting from the 1D
hydrodynamic expression \cite{Capuzzi2006}
\begin{equation}
c_1=\left\{\left.\frac{1}{M}\int d^2r_\perp \,n \right/ \int
d^2r_\perp \, \left[\frac{\partial \mu(n)}{\partial n}\right]^{-1}
\right\}^{1/2} ,\label{eq:c1_capuzzi}
\end{equation}
where the the density profile along the radial direction should be
evaluated in LDA. The above results hold for sound waves
characterized by wavelengths significantly larger than the radial
size of the gas. Carrying out a perturbative development around
unitarity, similar to the one employed above for the calculation of
the excitation frequencies, we obtain the result
\begin{equation}
\frac{\delta c_1}{c_1} = -\frac{3\zeta}{20\xi}\left(k_F
a\right)^{-1} ,\label{eq:delta_c1}
\end{equation}
where $c_1 =(\xi/5)^{1/2}(\hbar k_F/M)$ is the 1D sound velocity
depending on the density in the center of the trap via $k_F$. Notice
that $c_1$ differs from the sound velocity calculated at the central
value of the density by the factor $\sqrt{3/5}$. In
Fig.\ref{fig:Thomas} we show the experimental values of $c_1$ in the
unitary regime measured in \cite{Joseph2007} together with the slope
evaluated from Eq.(\ref{eq:delta_c1}) using for $\xi$ and $\zeta$
the $T\simeq 0$ values obtained from the experiment
\cite{Navon2010}. A quadratic fit is applied to the experiment data
(dashed red curve) \cite{Thomas_private}. The figure shows that  the
value of the slope at unitarity [black curve, $\delta c_1/c_1 \sim
-0.27/(k_F^0 a)$] overestimates the experimental linear changes in a
visible way, suggesting that these experimental data were carried
out at relatively higher temperatures, such that they cannot be
accurately reproduced by employing the $T=0$ values of the  contact
parameter. Another source of disagreement might be due to the fact
that the conditions of applicability of the 1D expression
(\ref{eq:c1_capuzzi}) for the sound velocity are not fully satisfied
in the experiment of \cite{Joseph2007}.

\section{Conclusion}

In conclusion our analysis reveals consistency between the
experimental results for the contact, obtained through the
measurement of the equation of state carried out at $T\simeq0$ in
\cite{Navon2010}, and the behavior of the collective frequencies
carried out in \cite{Altmeyer2007} at the lowest temperatures. It
would be interesting to extend our analysis of the frequency shifts
to finite temperature. The analysis could be simplified by the fact
that the scaling modes of monopole and quadrupole types have a
universal behavior at unitarity and their frequencies, calculated in
the hydrodynamic regime, are independent of temperature. Furthermore
they are  not coupled to second sound. The proper calculation of the
resulting slope at finite temperature and the explicit connection
with Tan's contact parameter at finite temperature will be the
object of a future work.

\acknowledgments

Useful discussions with R. Combescot and L. P. Pitaevskii are
acknowledged. This work has been supported by ERC through the QGBE
grant.


\begin{thebibliography}{99}
\bibitem{Note_02}
This follows from the fact that the unperturbed frequencies [see for
example Eq.(\ref{eq:dispersion_law})] do not depend on the value of
the chemical potential $\bar{\mu}$.
\end{thebibliography}
\end{document}